\documentclass[aps,pre,preprint,groupedaddress,eqsecnum,showpacs,longbibliography]{revtex4-1}
\usepackage{graphicx}
\usepackage{amsmath}

\begin{document}


\title{Motion of an optically torqued nanorod: the overdamped case}


\author{W. C. Kerr}
\email[]{wck@wfu.edu}
\affiliation{Olin Physical Laboratory, Wake Forest University, Winston-Salem, NC 27109}
\author{H. Nasif}
\affiliation{Department of Mathematics, Wake Forest University, Winston-Salem, NC 27109}
\author{S. Raynor}
\email[]{raynorsg@wfu.edu}
\affiliation{Department of Mathematics, Wake Forest University, Winston-Salem, NC 27109}


\date{\today}

\begin{abstract}
A recent experiment [W. A. Shelton {\emph{et\ al.}}, Phys.\ Rev.\ E {\bf{71}}, 036204 (2005)] measured the response of a nanorod trapped in a viscous fluid to the torque produced by an incident optical frequency electromagnetic wave.  The nonlinear differential equation describing this motion is similar that of a damped, driven pendulum.  The overdamped limit of this equation has been solved analytically.  We analyze the properties of this solution in comparison to the observations of the experiment and find very close agreement.
\end{abstract}


\maketitle

\section{\label{sec:Intro} Introduction}
Experimental study of the motion of nanoscale objects in fluids has been a topic of considerable interest in recent years~\cite{BoninKourmanovWalker:2002, SheltonBoninWalker:2005,Helgesen:1990,Keshoju:2007,Pedaci:2011}.  There is a considerable range in the parameters of the systems that can be constructed for study, including the relative magnitudes of the viscous forces versus the inertial forces.  These systems are also typically subject to external fields whose magnitudes and rates of variation can be varied.  Finally, the presence of surfaces near the nano-object can markedly change the motion.  Achieving a theoretical understanding of the resulting data presents interesting problems in nonlinear science.

The theoretical study reported here grows out of experiments employing optical torques applied to nanoscale objects, performed by the Bonin group~\cite{BoninKourmanovWalker:2002, SheltonBoninWalker:2005}.  Their system consisted of a nanoscale, anisotropically polarizable dielectric rod, held in a viscous medium by an optical trap, and torqued by a rotating electromagnetic field oscillating at an optical frequency.  The first experiment~\cite{BoninKourmanovWalker:2002} was complicated by the nearness of the nanorod to the walls of the apparatus.  The second experiment~\cite{SheltonBoninWalker:2005} employed a method to isolate the nanorod from the walls.  Newton's Second Law for rotation gives the equation of motion for the nanorod.  The net torque is a sum of the contributions from the torque produced by the incident electromagnetic wave acting on the induced electric dipole moment of the rod and from the frictional torque of the viscous medium.  Since the rate of oscillation of the electromagnetic field is many orders of magnitude more rapid than the motion of the rod, this elementary equation of motion was averaged over many of these oscillations to obtain an equation involving only slowly varying quantities.  For the parameters of the particular experimental system of interest~\cite{SheltonBoninWalker:2005}, the damping forces were much greater than the inertial forces, so the inertial term was neglected.  The resulting first-order nonlinear ordinary differential equation has been solved analytically.  We present this solution and apply it to the experiment of Ref.~\cite{SheltonBoninWalker:2005}.

There have been previous experimental~\cite{Helgesen:1990,Keshoju:2007} and theoretical~\cite{Ghosh:2013} investigations of the driven rotation of a nanoscale \emph{magnetic} object with a fixed permanent magnetic moment torqued by a rotating \emph{magnetic} field.  These descriptions of its motion indicate that it is very similar to that obtained for the system investigated here, {\emph{viz.}} a polarizable dielectric object driven by an electric field.

In the following, Sec.\ \ref{sec:Desc} describes the system,  Sec.\ \ref{sec:Solns} gives the solution of the overdamped equation of motion, describes its properties, and compares it to the experimental data, and  Sec.\ \ref{sec:Discussion} summarizes our results.

\section{\label{sec:Desc} Description}
The experimental arrangement is described in detail by Shelton {\textit{et al.}}~\cite{SheltonBoninWalker:2005}, and the reader is referred to that paper for figures showing the configuration.  (By clicking on the online link included in Ref.~\cite{SheltonBoninWalker:2005} and navigating to Fig.\ 3 on that web page, the reader can see the figure corresponding to the description here.)  The system consists of a cylinder (the nanorod) made from insulating material (borosilicate glass), held in an optical trap in a viscous medium (water), and driven by an incident linearly polarized optical beam.  Its diameter is approximately one-half micron and its length is a few microns.  The design of the optical trap ensures that the nanorod aligns with its long axis perpendicular to the propagation direction of the optical beam, so that the rod's motion is confined to a plane.  The electric field ${\bf{E}}$ of the optical beam induces a dipole moment ${\bf{p}}$ in the nanorod according to
${\bf{p}} = \mathord{\buildrel{\lower3pt\hbox{$\scriptscriptstyle\leftrightarrow$}}
\over \alpha }  \cdot {\bf{E}}$, which results in a torque
${\boldsymbol{\tau }} = {\bf{p}} \times {\bf{E}}$.
The tensor polarizability
$\mathord{\buildrel{\lower3pt\hbox{$\scriptscriptstyle\leftrightarrow$}}
\over \alpha }$
describes the anisotropy caused by shape birefringence~\cite{Stratton:1941, Hulst:1981, LanLifPit:1984}, which produces a moment $\bf{p}$ that is not parallel to $\bf{E}$.
Although the beam is nominally linearly polarized, its plane of polarization is rotated by passing it through a mechanically rotated half-wave plate; the polarization plane rotation rate $\Omega$ is very slow compared to the optical frequency $\omega$.  The torque created by the anisotropic polarizability and the rotation of the polarization of the light beam causes the nanorod to rotate in a plane perpendicular to the beam's propagation direction.  Newton's Second Law for rotation then gives for the equation of motion (EOM) of this system.  However, as just described, this EOM contains the very rapidly oscillating electric field and the much more slowly moving nanorod.  Therefore, it is appropriate to average out the rapidly varying part, which gives an ordinary differential equation (ODE) for only the slowly changing variable.  The details of this procedure are described in Appendix A of Ref.~\cite{SheltonBoninWalker:2005}.  The result is the ODE
\begin{equation}
I\frac{{{d^2}\theta }}{{d{t^2}}} = N\sin \left[ {2\left( {\Omega t - \theta } \right)} \right] - \gamma \frac{{d\theta }}{{dt}}.
     \label{eq:basicEOM}
\end{equation}
Here $\theta$ is the angle of rotation angle of the nanorod with respect to some fixed axis; we will refer to it as the physical angle.  $I$ is the moment of inertia of the nanorod, $N$ is the torque amplitude, $\Omega$ is the angular velocity of the polarization plane of the optical beam, and $\gamma$ is the rotational drag coefficient of the viscous medium.  The torque amplitude is determined by the amplitude of the electric field of the light beam $E_0$ and by the difference of the principal values of the polarizability parallel and perpendicular to the axis of the nanorod:
$N = {\textstyle{1 \over 4}}\left( {{\alpha _\parallel } - {\alpha _ \bot }} \right)E_0^2$.

Typical values for the parameters are also given in Ref.~\cite{SheltonBoninWalker:2005}:
$I = 3 \times {10^{ - 28}}{\ \rm{ kg}} \cdot {{\rm{m}}^2}$,
$\gamma  = 3.8 \times {10^{ - 20}}{\ \rm{ kg}} \cdot {{\rm{m}}^2}/{\rm{s}}$,
$N = 1.14 \times {10^{ - 18}}{\ \rm{ kg}} \cdot {{\rm{m}}^{\rm{2}}}{\rm{/}}{{\rm{s}}^{\rm{2}}}$.
The value of $\Omega$, the angular rotation frequency of the linear polarization, can be varied in the range of 3 - 300 rad/s.  The angular vibrational frequency of the optical beam is
${\Omega _{{\rm{opt}}}} = 3.67 \times {10^{15}}{\rm{\  rad/s}}$.  This value is much larger than the angular rotation frequency of the polarization plane of the optical beam, so the beam is truly slowly rotating linearly polarized light, not circularly polarized light.  As already stated above, this very high frequency has been integrated out to obtain the EOM in Eq.\ (\ref{eq:basicEOM}).

\section{\label{sec:Solns} Solutions}
The combination of variables appearing in Eq.\ (\ref{eq:basicEOM}) motivates introducing a new dependent variable
\begin{equation}
\phi = 2(\Omega t - \theta),
     \label{eq:auxAngle}
\end{equation}
which we will refer to as the auxiliary angle.
The EOM Eq.\ (\ref{eq:basicEOM}) in terms of this variable is
\begin{equation}
I \frac{{{d^2}\phi }}{{d{t^2}}} + \gamma \frac{{d\phi }}{{dt}} + 2N\sin \phi  = 2\gamma \Omega .
     \label{eq:auxAngleEOM}
\end{equation}

Next we ask if there are any stationary solutions of Eq.\ (\ref{eq:auxAngleEOM}), that is constant solutions with
$\dot{\phi} =  \ddot{\phi} = 0$.  Such solutions must satisfy
\begin{equation}
\sin \phi = \frac{\gamma \Omega}{N} .
     \label{eq:statSoln}
\end{equation}
For these stationary solutions, one sees from Eq.\ (\ref{eq:auxAngle}) that such solutions describe uniform rotation of the physical angle $\theta$ with the same angular velocity as the polarization of the optical field.  Clearly such solutions exist when the parameters are such that
$0< \gamma \Omega/N <1$, and in fact there are two solutions of Eq.\ (\ref{eq:statSoln}), with $\phi$ in the range $0 < \phi < \pi$.  A stability analysis of these two solutions, based on Eq.\ (\ref{eq:auxAngleEOM}), shows that the one in the range $0 < \phi < \pi/2$ is stable, and the one in the range $\pi/2 < \phi < \pi$ is unstable~\cite{StabilityNote}.

This search for stationary solutions has identified an intrinsic frequency scale
\begin{equation}
\Omega_c = \frac{N}{\gamma} = 30 {\rm{\  rad/s}} ,
     \label{eq:criticalOmega}
\end{equation}
where the value for $\Omega_c$ is obtained from the typical values of the parameters of the experiment listed in Sec.\ \ref{sec:Desc}.  We use $\Omega_c$ to introduce a dimensionless time as a new independent variable,
\begin{equation}
u = 2 \Omega_c t .
     \label{eq:dimlessTime}
\end{equation}
In terms of this new variable, the EOM in Eq.\ (\ref{eq:auxAngleEOM}) becomes
\begin{equation}
\epsilon \frac{d^2 \phi}{d u^2} + \frac{d \phi}{du} + \sin \phi = r ,
     \label{eq:rewrittenEOM}
\end{equation}
where the two dimensionless parameters are
\begin{equation}
r = \frac{\Omega}{\Omega_c}, \ \ \ \ \ \epsilon = \frac{2NI}{\gamma^2} =  \frac{2 I \Omega_c}{\gamma}.
     \label{eq:defEpsilon}
\end{equation}
The ratio $r$ is the control parameter, the ratio of the experimentally controllable angular rotation velocity of the optical polarization to the critical value from Eq.\ (\ref{eq:criticalOmega}).  In terms of $r$, the stable stationary solution of Eq.\ (\ref{eq:statSoln}) is
\begin{equation}
\phi = \sin^{-1} (r),
     \label{eq:stableStatSoln}
\end{equation}
with the customary understanding of the principal value range of the inverse sine function.  Using the typical values from Sec.\ \ref{sec:Desc} the parameter $\epsilon$ is about $4.7 \times 10^{-7}$; this very small value justifies identifying $\epsilon$ as the perturbation parameter and dropping the first (inertial) term from Eq.\ (\ref{eq:rewrittenEOM}).
Thus we base the subsequent analysis on the first-order nonlinear inhomogeneous ordinary differential equation,
\begin{equation}
\frac{d \phi}{du} + \sin \phi = r  .
     \label{eq:auxEOM}
\end{equation}

Equation (\ref{eq:auxEOM}) has the same form as the dimensionless EOM of an overdamped pendulum in a gravitational field and driven by a constant torque.  In his monograph~\cite{Strogatz:1994} Strogatz presented a qualitative analysis of it that illustrates the salient features. Previously Adler~\cite{Adler:1946} gave the analytic solution, in the context of an electronic oscillator circuit.  We describe his solution in our notation and adapt it to the details of the experiment discussed here~\cite{SheltonBoninWalker:2005}.

The ODE in Eq.\ (\ref{eq:auxEOM}) is separable and can be formally integrated as
\begin{equation}
     \int_0^{\phi}  \frac{d \phi^\prime}{r - \sin \phi^\prime} = \int_0^{u} du^\prime = u .
          \label{eq:quadrature}
\end{equation}
In writing Eq.\ (\ref{eq:quadrature}), we assumed the initial condition that at time zero, $u=0$, the nanorod has initial physical angle and also auxiliary angle equal to zero.  The experimental initial orientation is not known, so we are free to take the initial orientation of the nanorod to define the axis for measuring angles.  Thus the assumptions used in writing Eq.\ (\ref{eq:quadrature}) are without loss of generality.

The integral in Eq.\ (\ref{eq:quadrature}) was executed by Adler~\cite{Adler:1946} and nowadays can be performed using mathematical software packages.  The result can then be inverted to obtain a solution for the auxiliary angle, $\phi (u)$.  The result depends on whether the control parameter $r$ is less than or greater than unity, so we present these two cases in turn.

\subsection{Case $r < 1$}
For this parameter range the solution for the auxiliary angle is
\begin{equation}
\phi (u) = 2 \tan^{-1} \left\{\frac{1-\sqrt{1-r^2}\tanh\left[\frac{1}{2}\sqrt{1-r^2} \ u + \tanh^{-1}\left(\frac{1}{\sqrt{1-r^2}}\right)\right]}{r} \right\} .
     \label{eq:Final_rLT1}
\end{equation}

From Eq.\ (\ref{eq:Final_rLT1}), one sees that
\begin{equation}
\lim_{u \rightarrow \infty} \phi (u) = 2 \tan^{-1} \left\{ \frac{1-\sqrt{1-r^2}}{r} \right\} = \sin^{-1} (r),
    \label{eq:AsympValue}
\end{equation}
i.e. at large time this solution asymptotically approaches the stable stationary solution found in Eq.\ (\ref{eq:stableStatSoln}).
\begin{figure}[h]
\includegraphics*[width=3.375in]{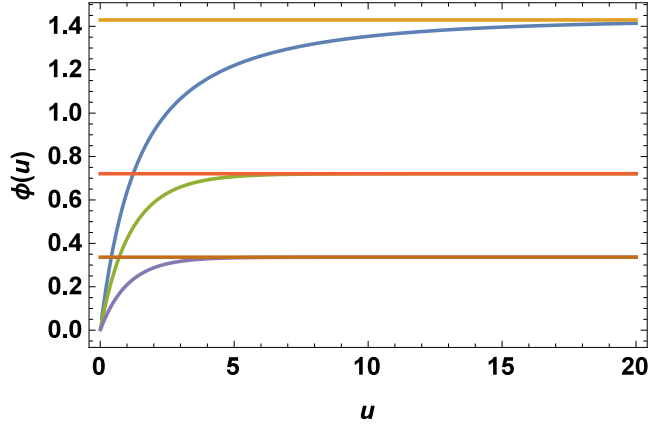}
\caption{(Color online) The auxiliary angle $\phi(u)$ versus $u$ for three values of $r$ smaller than unity.  From bottom to top, the values are $r=0.33,\ 0.66,\ 0.99$.  The horizontal lines are the asymptotic values $\sin^{-1} (r)$ for the corresponding $r$ values, from Eq.\ (\ref{eq:AsympValue}).}
\label{fig:AuxAngle3rValuesSmaller1}
\end{figure}
This solution for the auxiliary angle, for three different values of the control parameter $r$, is shown in Fig.\ \ref{fig:AuxAngle3rValuesSmaller1}.  For these solutions, at large $u$ the physical angle $\theta$ rotates uniformly at angular velocity $\Omega$.  Graphs of $\theta$ versus physical time $t$, from Eqs.\ (\ref{eq:auxAngle}) and (\ref{eq:dimlessTime}), would asymptotically become straight lines with slope $\Omega$ and with an $r$-dependent offset due to the approach of the auxiliary angle $\phi$ to its asymptotic value.

\begin{figure}[h]
\includegraphics*[width=5.375in]{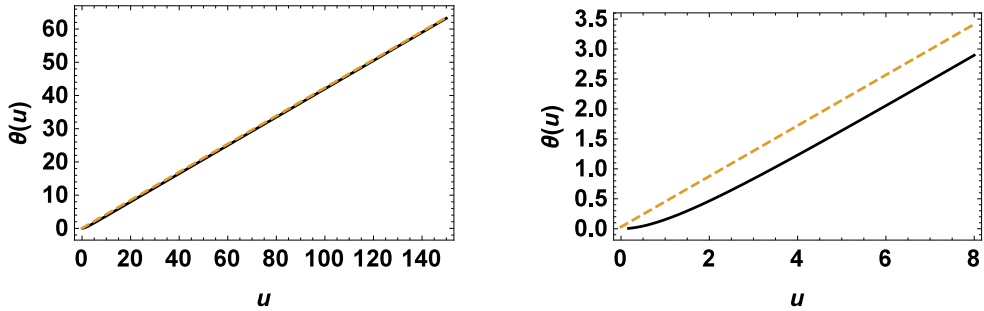}
\caption{(Color online) Comparison of the experimental fit to the theoretical result for the physical angle $\theta$ for $r = 0.85$.  The fit to the experimental data [Eq.\ (\ref{eq:ExpFit})] is the dotted lines, and the theoretical equation [Eq.\ (\ref{eq:Final_rLT1})] is the solid lines.  The left panel shows the comparison for the whole range of time measured by the experiment.  The right panel shows the comparison for short times.}
\label{fig:PhysAngleThAndExp}
\end{figure}

In Fig.\ 6 of Ref.~\cite{SheltonBoninWalker:2005}, those authors present a fit to their data for the physical angle $\theta$, for $r = 0.85$; it is
\begin{equation}
\theta_{\textrm{exp}} (u) = 0.423 u + 0.03
     \label{eq:ExpFit}
\end{equation}
(converted from their physical time scale $t$ in seconds to dimensionless time $u$).  In our Fig.\ \ref{fig:PhysAngleThAndExp} we compare the combination of our Eqs.\ (\ref{eq:auxAngle}) and (\ref{eq:Final_rLT1}) to their fit, Eq.\ (\ref{eq:ExpFit}).
The left panel shows the comparison over the full time range presented in Ref.~\cite{SheltonBoninWalker:2005}.  The experimental data and the theoretical equation agree quite closely.  The right panel shows the comparison of the short time behavior, near $u = 0$.  Equation (\ref{eq:ExpFit}) for the data is a linear function of $u$.  We perform a series expansion on Eq.\ (\ref{eq:Final_rLT1}) and combine it with Eq.\ (\ref{eq:auxAngle}) to find that the theoretical formula for the physical angle $\theta (u)$ is not linear but is initially quadratic,
\begin{equation}
\theta (u) = \frac{1}{4} r u^2 - \frac{1}{12} r u^3 + \cdots .
     \label{eq:shortTimeTheta}
\end{equation}
Thus for small times the experimental and theoretical descriptions differ, as shown in the right panel of Fig.\ \ref{fig:PhysAngleThAndExp}.

This analysis of the short time behavior is somewhat artificial.  The first-order ODE Eq.\ (\ref{eq:auxEOM}) allows only the specification of the initial angle, and the initial auxiliary angular velocity is then determined.  The combination of Eqs.\ (\ref{eq:auxEOM}) and (\ref{eq:auxAngle}) require the initial angular velocity of the \emph{physical} angle to be zero, as is also seen from Eq.\ (\ref{eq:shortTimeTheta}).  The more complete EOM, Eqs.\ (\ref{eq:basicEOM}) or (\ref{eq:auxAngleEOM}), being second-order ODEs, also require specification of the initial angular velocity, but this second initial condition can not be dealt with by Eq.\ (\ref{eq:auxEOM}). Therefore the lesson to be drawn from the right panel of Fig.\ \ref{fig:PhysAngleThAndExp} is to appreciate how rapidly the overdamped motion sets in and not to focus on the disagreement between the data and the theory for short time.

\subsection{Case $r > 1$}

For this parameter range, we see from Eq.\ (\ref{eq:auxEOM}) that when $r>1$, the derivative $d \phi / du$ is always positive, so $\phi (u)$ is a monotonically increasing function.  The solution of Eq.\ (\ref{eq:auxEOM})  that has this property is
\begin{equation}
\phi(u) = 2 \tan ^{-1} \left\{ \frac{1+\sqrt{r^2-1}\tan \left[ f(u) \right]} {r}\right\}
+ 2 \pi \lfloor \frac{f(u)}{\pi} + \frac{1}{2} \rfloor .
     \label{eq:AuxAngle_rLarger1}
\end{equation}
Here $\lfloor x \rfloor$ denotes the largest integer less than $x$, and the other function introduced here is
\begin{equation}
f(u) = \frac{1}{2}\sqrt{r^2-1} \  u - \tan^{-1}\left(\frac{1}{\sqrt{r^2-1}}\right)
     \label{eq:ArgOfTan}
\end{equation}

Standard differential equation techniques to solve Eq.\ (\ref{eq:auxEOM}) [or to integrate Eq.\ (\ref{eq:quadrature})] give just the first term of Eq.\ (\ref{eq:AuxAngle_rLarger1}), but that term alone is not sufficient to describe the motion of the nanorod for $r>1$.  First, the first term of Eq.\ (\ref{eq:AuxAngle_rLarger1}) is a periodic function of $u$ instead of being monotonically increasing.  Second, by convention the range of the inverse tangent function is restricted to the interval $( - \pi /2 , + \pi /2)$, but the actual range of the nanorod angle is unbounded.  Third, the first term of Eq.\ (\ref{eq:AuxAngle_rLarger1}) is discontinuous because the tangent function jumps from $+ \infty$ to $- \infty$ every time $f(u)$ passes an odd integer multiple of $\pi/2$ as $u$ increases; the actual motion of the nanorod is continuous.  All of these shortcomings of the first term of Eq.\ (\ref{eq:AuxAngle_rLarger1}) are overcome by adding the second term.  At each value of $u$ where the first term of Eq.\ (\ref{eq:AuxAngle_rLarger1}) has a discontinuity, the second term removes it by adding an additional $2 \pi$.  In this way we obtain a solution of Eq.\ (\ref{eq:auxEOM}) that is continuous and monotone increasing (as seen in the subsequent figures).

The analysis here somewhat mirrors that in Ref.~\cite{Adler:1946}, but the equivalent of Eq.\ (\ref{eq:AuxAngle_rLarger1}) does not appear in that paper. The reason may be that the variable $\phi$ in Ref.~\cite{Adler:1946} is the phase difference between two signals in an electronic circuit, so that only \emph{phase factors} e,g. $\cos \phi$ and $\sin \phi$ are important.  Thus the value of the phase can be considered modulo $ 2 \pi$.  In the experiment of Ref.~\cite{SheltonBoninWalker:2005}, the physical angle $\theta$ (and therefore the auxiliary angle $\phi$) is measured and is visually observed to perform many revolutions, so to compare with that data it is necessary here to calculate its actual value.

\begin{figure}[h]
\includegraphics*[width=5.375in]{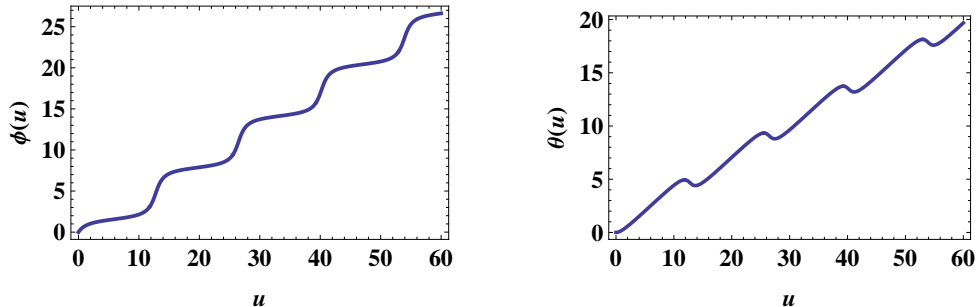}
\caption{(Color online) The auxiliary angle $\phi (u)$ (left panel) and the physical angle $\theta (u)$ (right panel) for $r=1.1$.  Equation (\ref{eq:auxAngle}) is the relation between the two angles.}
\label{fig:Angles_rEq1p1}
\end{figure}

Equation (\ref{eq:AuxAngle_rLarger1}) for the auxiliary angle is plotted in Fig.\ \ref{fig:Angles_rEq1p1} (left panel) for a particular value $r=1.1$.  The experiment measures the physical angle $\theta$, which is related to the auxiliary angle $\phi$ by Eq.\ (\ref{eq:auxAngle}); the physical angle is also shown in Fig.\ \ref{fig:Angles_rEq1p1} (right panel) for $r=1.1$.  The physical angle has ``flipbacks'' (so-called in Ref.\ ~\cite{SheltonBoninWalker:2005}), i.e. intervals where $d \theta /du$ is negative.

\begin{figure}[h]
\includegraphics*[width=5.375in]{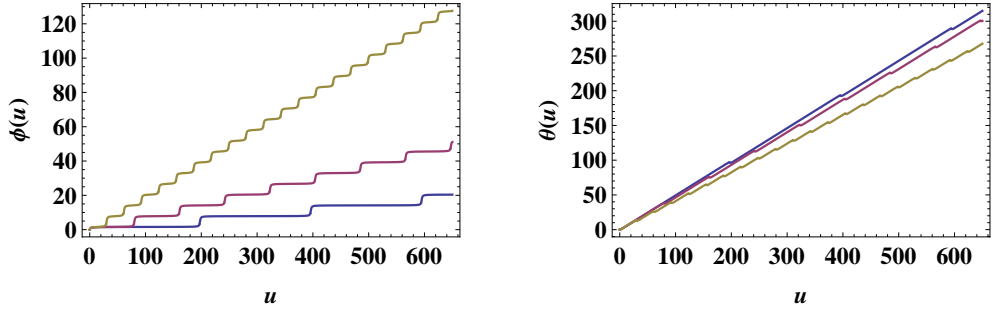}
\caption{(Color online)The auxiliary angle $\phi (u)$ (left panel) and the physical angle $\theta (u)$ (right panel) for three values of $r$ ($r=1.02, \ 1.003, \ 1.0005$) approaching unity from above, showing the increase in the period of the rapid angular changes for $r \rightarrow 1^+$.  In the left panel the $r$-values decrease towards $1$ going from the top curve to the bottom one.  In the right panel the sequence is reversed: $r$ decreases toward $1$ from the bottom curve to the top.}
\label{fig:AuxAndPhysAngles3rValuesBigger1}
\end{figure}

The bursts of rapid change of the auxiliary angle seen in the left half of Fig.\ \ref{fig:Angles_rEq1p1} are periodic, as described by the tangent function in Eq.\ (\ref{eq:AuxAngle_rLarger1}).  Since the period of the tangent function is $\pi$, the period $T_u$ (with respect to dimensionless time $u$) of the intervals between large velocity changes is
\begin{equation}
T_u = \frac{2 \pi}{\sqrt{r^2-1}} \rightarrow \frac{\sqrt{2} \pi}{\sqrt{r-1}} \ \ \ \mathrm{as} \ \ \  r \rightarrow 1^+ .
\end{equation}
The last step shows that this period increases without bound as $r \rightarrow 1$ from above with a square root singularity.  This increase in the time between the rapid changes is seen in Fig.\ \ref{fig:AuxAndPhysAngles3rValuesBigger1} (left panel), which plots $\phi (u)$ for three different $r$-values approaching unity from above.  The right panel of Fig. \ref{fig:AuxAndPhysAngles3rValuesBigger1} shows the corresponding curves for the physical angle $\theta (u)$.  One also sees here that the average angular velocity of the \emph{physical} angle i.e. $\langle {d \theta /du} \rangle$ increases for $r \rightarrow 1^+$, in qualitative agreement with the experiment (Fig.\ 9 of Ref.\ ~\cite{SheltonBoninWalker:2005} for $r>1$).

\begin{figure}[h]
\includegraphics*[width=3.375in]{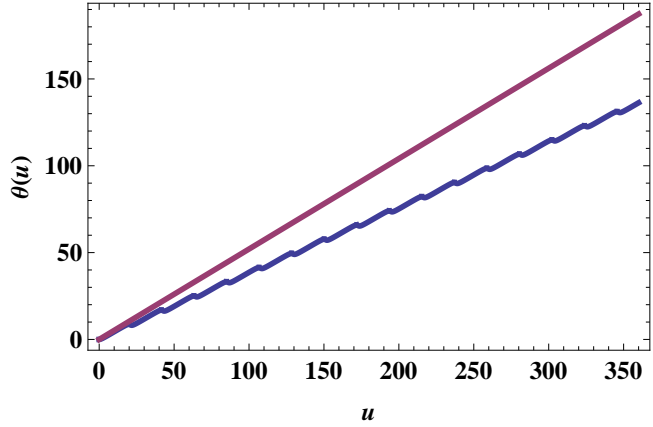}
\caption{(Color online) The polarization angle $\Omega t = r u/2$ (the straight line) and the physical angle $\theta (u)$ (the jagged curve) for $r=1.041$.
Ref.~\cite{SheltonBoninWalker:2005} has this same graph in their Fig.\ 7 with the experimental data.}.
\label{fig:PolarizationAndPhysAngles}
\end{figure}

Reference~\cite{SheltonBoninWalker:2005} includes a figure showing the physical angle $\theta (u)$ in comparison to the rotation angle of the light polarization, so we show the same comparison in Fig.\ \ref{fig:PolarizationAndPhysAngles}, for the same $r$-value, $r=1.041$.  The two figures are nearly the same, but in the experimental one the increase of $\theta (u)$ with time has intervals with different behaviour, due to experimental features such as variation in the rotation speed of the half-wave plate that is rotating the polarization plane of the light or variation in the experimental $\Omega_c$ due to drifts in properties of the experimental apparatus.  None of those things happens in the theoretical equations.

\subsection{\label{sec:Summary}Average Angular Velocities}
The authors of Ref.~\cite{SheltonBoninWalker:2005} summarized their data by plotting the average physical angular velocity
$\langle d \theta / dt \rangle$ as a function of the plane of polarization rotation rate $\Omega$.  Here we show that our analytical results in Eqs.\ (\ref{eq:Final_rLT1}) and (\ref{eq:AuxAngle_rLarger1}) provide a confirmation of their conclusions.

The average rotation rates of the physical angle and the auxiliary angle are related by the average of the derivative of Eq. (\ref{eq:auxAngle}),
\begin{equation}
\left\langle \frac{d \theta}{du} \right\rangle = \frac{1}{2} r - \frac{1}{2} \left\langle \frac{d \phi}{du} \right\rangle .
     \label{eq:avgPhysAndAuxAngles}
\end{equation}
The average value of $d \phi /du$ over an interval $(u_1, u_2)$ is obtained from
\begin{equation}
\left\langle \frac{d \phi}{du} \right\rangle = \frac{1}{u_2 - u_1} \int_{u_1}^{u_2} \frac{d \phi}{d u^\prime} d u^\prime
  = \frac{\phi (u_2) - \phi (u_1)}{u_2 - u_1}.
       \label{eq:AvgAuxAngVel}
\end{equation}
The evaluation of the last expression in Eq.\ (\ref{eq:AvgAuxAngVel}) depends on the value of $r$.

\subsubsection{$r < 1$}
For this case $\phi (u)$ is monotonic [Fig.\ \ref{fig:AuxAngle3rValuesSmaller1}], so we take the interval to be from $u_1 = 0$ to $u_2 = \infty$.  Then the numerator of Eq.\ (\ref{eq:AvgAuxAngVel}) is the value in Eq.\ (\ref{eq:AsympValue}), and the denominator is infinite, so
\begin{equation}
\left\langle \frac{d \phi}{du} \right\rangle = 0.
\end{equation}
From Eq.\ (\ref{eq:avgPhysAndAuxAngles}), $\langle d \theta / du \rangle = r/2$, which, when converted to physical time $t$, gives
\begin{equation}
\left\langle \frac{d \theta}{dt} \right\rangle = \Omega , \ \ \ \ \ \Omega < \Omega_c .
\end{equation}
This result confirms both the result in Ref.~\cite{SheltonBoninWalker:2005} and our statement after Eq.\ (\ref{eq:AsympValue}) that graphs of $\theta (t)$ are asymptotic to straight lines with slope $\Omega$.

\subsubsection{$r > 1$}
In this range of $r$, we recognize from Fig.\ \ref{fig:Angles_rEq1p1} that the average is the same over any interval between two successive points where $f(u)$ [Eq.\ (\ref{eq:ArgOfTan})] equals an odd multiple of $\pi /2$, for example the two points where $f(u_1) = \pi /2$ and $f(u_2) = 3 \pi /2$.  From Eq.\ (\ref{eq:ArgOfTan}) we determine the denominator of Eq.\ (\ref{eq:AvgAuxAngVel}) to be $u_2 - u_1 = 2 \pi / \sqrt{r^2 - 1}$.  To obtain the numerator of Eq.\ (\ref{eq:AvgAuxAngVel}) we use Eq.\ (\ref{eq:AuxAngle_rLarger1}).  The values of $u_1$ and $u_2$ are points where the two terms of Eq.\ (\ref{eq:AuxAngle_rLarger1}) have cancelling discontinuities, so the evaluation is simplest if we take $u_1$ and $u_2$ to approach the endpoints of the interval from its interior.  The result is $\phi (u_2) - \phi (u_1) = 2 \pi$.  We substitute these values into Eq.\ (\ref{eq:AvgAuxAngVel}) and obtain
\begin{equation}
\left\langle \frac{d \phi}{du} \right\rangle = \sqrt{r^2 - 1} .
\end{equation}
We convert this result to the physical angle $\theta$ and physical time $t$, and obtain
\begin{equation}
\left\langle \frac{d \theta}{dt} \right\rangle = \Omega - \sqrt{\Omega^2 - \Omega_c^2} , \ \ \ \ \ \ \ \ \Omega > \Omega_c .
\end{equation}
This equation shows that the average physical angular velocity increases as $\Omega$ approaches $\Omega_c$ from above, as is shown in the right panel of Fig.\ \ref{fig:AuxAndPhysAngles3rValuesBigger1}.  It also agrees with the result of Ref.~\cite{SheltonBoninWalker:2005}.

\section{\label{sec:Discussion} Discussion}
In this paper we have compared the analytic solution of the EOM for an overdamped, driven nanorod to its measured motion.  In agreement with the experiment, the analytic solution identifies two regimes with qualitatively different motions.  The distinction is based on the ratio of the rotation frequency of the polarization plane of the incident light providing the torque to a frequency scale intrinsic to the system.  When this ratio is smaller than one, the rod rotation monotonically approaches uniform rotation following the polarization of the light field.  When the ratio is larger than one, the rod rotation is non-uniform and exhibits ``flipbacks'', where the angular velocity reverses for short time intervals.  One feature, not apparent in the experimental data but which comes out of the analytic solution, is that the time interval between flipbacks diverges as this ratio approaches unity from above.

In Sec.\ \ref{sec:Solns} we introduced a frequency scale $\Omega_c$ with a corresponding time-scale
\begin{equation}
t_B = \frac{\gamma}{2 N} = 1.67 \times 10^{-2} {\rm{\  sec}} ,
     \label{eq:BoninScale}
\end{equation}
in terms of which the dimensionless time is $u = t/t_B$.  The ODE for the auxiliary angle $\phi$ using this time scale is Eq.\ (\ref{eq:rewrittenEOM}), in which a perturbation parameter $\epsilon$ is identified [Eq.\ \ref{eq:defEpsilon})].  It is small when evaluated with the parameter values in Sec.\ \ref{sec:Desc}.  The solution we have obtained in Sec.\ \ref{sec:Solns} is the $\epsilon = 0$ or zeroth-order (in perturbation theory) solution.  Since $\epsilon$ appears in the highest-order derivative term in Eq.\ (\ref{eq:rewrittenEOM}), that term is a singular perturbation~\cite{BenderOrszag:1978}, e.g. the full second-order ODE has more linearly-independent solutions than the first-order ODE for $\epsilon = 0$ (be careful to distinguish between the order of the ODE and the order in perturbation theory).  This formulation has the advantage that the zeroth-order (in perturbation theory) solution correctly identifies the qualitatively different types of motion that occur when the control parameter is less than or greater than unity, in agreement with the experiment.

Another combination of the parameters in Eq.\ (\ref{eq:basicEOM}) with dimension of time is
\begin{equation}
t_K = \frac{I}{\gamma} = 7.89 \times 10^{-9} {\rm{\ sec}} ,
     \label{eq:KerrTimeScale}
\end{equation}
when evaluated using the parameter values in Sec. \ref{sec:Desc}.
This scale is much smaller than the $t_B$ scale and is inaccessible by the experiment described in Ref.~\cite{SheltonBoninWalker:2005}.  If we use this scale to define a different dimensionless time
\begin{equation}
s = \frac{t}{t_K} ,
\end{equation}
then the ODE for the auxiliary angle is
\begin{equation}
\frac{d^2 \phi}{ds^2} + \frac{d \phi}{ds} + \epsilon (\sin \phi - r) = 0 .
     \label{eq:KerrScaleODE}
\end{equation}
Here the perturbation is the nonlinear term, instead of the highest-order derivative term, and the $\epsilon = 0$ equation (zeroth-order in perturbation theory) is linear.  With the same initial condition employed in Eq.\ (\ref{eq:quadrature}), the solution of the $\epsilon = 0$ equation is
\begin{equation}
\phi^{(0)} (s) = \omega_0 (1 - e^{-s} ) ,
\end{equation}
where the superscript $(0)$ denotes the order in perturbation theory and $\omega_0$ is the initial value $d \phi / ds |_{s=0}$.  This solution is qualitatively similar to the $r<1$ case only of the solution obtained in Sec.\ \ref{sec:Solns}.  Uncovering the distinction between the $r<1$ and $r>1$ cases would have to emerge from higher orders in perturbation theory using this formulation of the problem.

One can observe that
\begin{equation}
\epsilon = \frac{t_K}{t_B} ,
\end{equation}
so that the disparity of these time scales is equivalent to $\epsilon$ having a small value, when these constants are evaluated with the system values given in Sec.\ \ref{sec:Desc}.  One can speculate that the motion might be much different if the system parameters were such that the two time scales $t_K$ and $t_B$ were more nearly comparable, which could be achieved with changes such as smaller damping, larger moment of inertia, or larger torque amplitude.  However, if one is searching for more complicated motions, it is useful to rewrite the second-order ODE Eq.\ (\ref{eq:auxAngleEOM}) as a set of first-order ODEs.  Since Eq.\ (\ref{eq:auxAngleEOM}) is autonomous, there are only two equations, which are
\begin{equation}
\omega (t) = \frac{d \phi (t)}{dt} , \ \ \ \ \  I \frac{d \omega (t)}{dt} = 2 \gamma \Omega - \gamma \omega (t) - 2N \sin[\phi (t)] .
\end{equation}
These equations identify that the phase space of this system is two-dimensional, which has the consequence that the  Poincar\'{e}-Bendixson~\cite{LichtLieb:1983} theorem says that chaotic motion is impossible.

Since the publication of~\cite{SheltonBoninWalker:2005}, further work related to it has been done.  In addition to~\cite{Keshoju:2007}, we list several several papers that utilize the optical trapping and/or optical torquing techniques that are used in the experiment of~\cite{SheltonBoninWalker:2005}.  On the theoretical side more elaborate calculations have been performed of the forces and torques exerted by an electromagnetic field with various polarization states on objects of different shapes and polarizabilities~\cite{Bonessi:2007,LiawKuo:2014,TrojekZemanek:2012}.  On the experimental side varied work has been done.  Diamond nano-crystals with nitrogen-vacancy defects have been held by optical traps and their ESR spectra studied to understand the dynamics of the nanocrystals in a viscous medium~\cite{HorowitzAwschalom:2012}.  A system similar to that of~\cite{SheltonBoninWalker:2005} has been studied in \cite{PedaciDekker:2011}; this experiment also includes an extensive analysis of the effects of random noise torques on the dynamics of the trapped object.  Finally, these techniques of optical trapping and optical torquing have been applied to the study of molecular motions in biological systems~\cite{LipfertDekker:2015}.

\begin{acknowledgements}
The authors thank Professor Keith D. Bonin for helpful discussions. H.\ N.\ thanks the Undergraduate Research and Creative Activities Center of Wake Forest College for a summer fellowship that supported this work.
\end{acknowledgements}


\begin{thebibliography}{20}%
\makeatletter
\providecommand \@ifxundefined [1]{%
 \@ifx{#1\undefined}
}%
\providecommand \@ifnum [1]{%
 \ifnum #1\expandafter \@firstoftwo
 \else \expandafter \@secondoftwo
 \fi
}%
\providecommand \@ifx [1]{%
 \ifx #1\expandafter \@firstoftwo
 \else \expandafter \@secondoftwo
 \fi
}%
\providecommand \natexlab [1]{#1}%
\providecommand \enquote  [1]{``#1''}%
\providecommand \bibnamefont  [1]{#1}%
\providecommand \bibfnamefont [1]{#1}%
\providecommand \citenamefont [1]{#1}%
\providecommand \href@noop [0]{\@secondoftwo}%
\providecommand \href [0]{\begingroup \@sanitize@url \@href}%
\providecommand \@href[1]{\@@startlink{#1}\@@href}%
\providecommand \@@href[1]{\endgroup#1\@@endlink}%
\providecommand \@sanitize@url [0]{\catcode `\\12\catcode `\$12\catcode
  `\&12\catcode `\#12\catcode `\^12\catcode `\_12\catcode `\%12\relax}%
\providecommand \@@startlink[1]{}%
\providecommand \@@endlink[0]{}%
\providecommand \url  [0]{\begingroup\@sanitize@url \@url }%
\providecommand \@url [1]{\endgroup\@href {#1}{\urlprefix }}%
\providecommand \urlprefix  [0]{URL }%
\providecommand \Eprint [0]{\href }%
\providecommand \doibase [0]{http://dx.doi.org/}%
\providecommand \selectlanguage [0]{\@gobble}%
\providecommand \bibinfo  [0]{\@secondoftwo}%
\providecommand \bibfield  [0]{\@secondoftwo}%
\providecommand \translation [1]{[#1]}%
\providecommand \BibitemOpen [0]{}%
\providecommand \bibitemStop [0]{}%
\providecommand \bibitemNoStop [0]{.\EOS\space}%
\providecommand \EOS [0]{\spacefactor3000\relax}%
\providecommand \BibitemShut  [1]{\csname bibitem#1\endcsname}%
\let\auto@bib@innerbib\@empty
\bibitem [{\citenamefont {Bonin}\ \emph {et~al.}(2002)\citenamefont {Bonin},
  \citenamefont {Kourmanov},\ and\ \citenamefont
  {Walker}}]{BoninKourmanovWalker:2002}%
  \BibitemOpen
  \bibfield  {author} {\bibinfo {author} {\bibfnamefont {K.~D.}\ \bibnamefont
  {Bonin}}, \bibinfo {author} {\bibfnamefont {B.}~\bibnamefont {Kourmanov}}, \
  and\ \bibinfo {author} {\bibfnamefont {T.~G.}\ \bibnamefont {Walker}},\
  }\bibfield  {title} {\enquote {\bibinfo {title} {Light torque nanocontrol,
  nanomotors and nanorockers},}\ }\href
  {http://www.opticsexpress.org/abstract.cfm?URI=oe-10-19-984} {\bibfield
  {journal} {\bibinfo  {journal} {Opt. Express}\ }\textbf {\bibinfo {volume}
  {10}},\ \bibinfo {pages} {984--989} (\bibinfo {year} {2002})}\BibitemShut
  {NoStop}%
\bibitem [{\citenamefont {Shelton}\ \emph {et~al.}(2005)\citenamefont
  {Shelton}, \citenamefont {Bonin},\ and\ \citenamefont
  {Walker}}]{SheltonBoninWalker:2005}%
  \BibitemOpen
  \bibfield  {author} {\bibinfo {author} {\bibfnamefont {W.~A.}\ \bibnamefont
  {Shelton}}, \bibinfo {author} {\bibfnamefont {K.~D.}\ \bibnamefont {Bonin}},
  \ and\ \bibinfo {author} {\bibfnamefont {T.~G.}\ \bibnamefont {Walker}},\
  }\bibfield  {title} {\enquote {\bibinfo {title} {Nonlinear motion of
  optically torqued nanorods},}\ }\href {\doibase 10.1103/PhysRevE.71.036204}
  {\bibfield  {journal} {\bibinfo  {journal} {Phys. Rev. E}\ }\textbf {\bibinfo
  {volume} {71}},\ \bibinfo {pages} {036204} (\bibinfo {year}
  {2005})}\BibitemShut {NoStop}%
\bibitem [{\citenamefont {Helgesen}\ \emph {et~al.}(1990)\citenamefont
  {Helgesen}, \citenamefont {Pieranski},\ and\ \citenamefont
  {Skjeltorp}}]{Helgesen:1990}%
  \BibitemOpen
  \bibfield  {author} {\bibinfo {author} {\bibfnamefont {G.}~\bibnamefont
  {Helgesen}}, \bibinfo {author} {\bibfnamefont {P.}~\bibnamefont {Pieranski}},
  \ and\ \bibinfo {author} {\bibfnamefont {A.T.}\ \bibnamefont {Skjeltorp}},\
  }\bibfield  {title} {\enquote {\bibinfo {title} {Nonlinear phenomena in
  systems of magnetic holes},}\ }\href {\doibase
  http://dx.doi.org/10.1103/PhysRevLett.64.1425} {\bibfield  {journal}
  {\bibinfo  {journal} {Physical Review Letters}\ }\textbf {\bibinfo {volume}
  {64}},\ \bibinfo {pages} {1425--1428} (\bibinfo {year} {1990})}\BibitemShut
  {NoStop}%
\bibitem [{\citenamefont {Keshoju}\ \emph {et~al.}(2007)\citenamefont
  {Keshoju}, \citenamefont {Xing},\ and\ \citenamefont {Sun}}]{Keshoju:2007}%
  \BibitemOpen
  \bibfield  {author} {\bibinfo {author} {\bibfnamefont {K.}~\bibnamefont
  {Keshoju}}, \bibinfo {author} {\bibfnamefont {H.}~\bibnamefont {Xing}}, \
  and\ \bibinfo {author} {\bibfnamefont {L.}~\bibnamefont {Sun}},\ }\bibfield
  {title} {\enquote {\bibinfo {title} {Magnetic field driven nanowire rotation
  in suspension},}\ }\href@noop {} {\bibfield  {journal} {\bibinfo  {journal}
  {Applied Physics Letters}\ }\textbf {\bibinfo {volume} {91}},\ \bibinfo
  {pages} {123114} (\bibinfo {year} {2007})}\BibitemShut {NoStop}%
\bibitem [{\citenamefont {Pedaci}\ \emph
  {et~al.}(2011{\natexlab{a}})\citenamefont {Pedaci}, \citenamefont {Huang},
  \citenamefont {van Oene}, \citenamefont {Barland},\ and\ \citenamefont
  {Dekker}}]{Pedaci:2011}%
  \BibitemOpen
  \bibfield  {author} {\bibinfo {author} {\bibfnamefont {F.}~\bibnamefont
  {Pedaci}}, \bibinfo {author} {\bibfnamefont {Z.}~\bibnamefont {Huang}},
  \bibinfo {author} {\bibfnamefont {Maarten}\ \bibnamefont {van Oene}},
  \bibinfo {author} {\bibfnamefont {S.}~\bibnamefont {Barland}}, \ and\
  \bibinfo {author} {\bibfnamefont {N.H.}\ \bibnamefont {Dekker}},\ }\bibfield
  {title} {\enquote {\bibinfo {title} {Excitable particles in an optical torque
  wrench},}\ }\href {\doibase doi:10.1038/nphys1862} {\bibfield  {journal}
  {\bibinfo  {journal} {Nature Physics}\ }\textbf {\bibinfo {volume} {7}},\
  \bibinfo {pages} {259--264} (\bibinfo {year}
  {2011}{\natexlab{a}})}\BibitemShut {NoStop}%
\bibitem [{\citenamefont {Ghosh}\ \emph {et~al.}(2013)\citenamefont {Ghosh},
  \citenamefont {Mandal}, \citenamefont {Karmakar},\ and\ \citenamefont
  {Ghosh}}]{Ghosh:2013}%
  \BibitemOpen
  \bibfield  {author} {\bibinfo {author} {\bibfnamefont {A.}~\bibnamefont
  {Ghosh}}, \bibinfo {author} {\bibfnamefont {P.}~\bibnamefont {Mandal}},
  \bibinfo {author} {\bibfnamefont {S.}~\bibnamefont {Karmakar}}, \ and\
  \bibinfo {author} {\bibfnamefont {A.}~\bibnamefont {Ghosh}},\ }\bibfield
  {title} {\enquote {\bibinfo {title} {Analytical theory and stability analysis
  of an elongated nanoscale object under external torque},}\ }\href {\doibase
  10.1039/c3cp50701g} {\bibfield  {journal} {\bibinfo  {journal} {Physical
  Chemistry Chemical Physics}\ }\textbf {\bibinfo {volume} {15}},\ \bibinfo
  {pages} {10817--10823} (\bibinfo {year} {2013})}\BibitemShut {NoStop}%
\bibitem [{\citenamefont {Stratton}(1941)}]{Stratton:1941}%
  \BibitemOpen
  \bibfield  {author} {\bibinfo {author} {\bibfnamefont {J.}~\bibnamefont
  {Stratton}},\ }\href@noop {} {\emph {\bibinfo {title} {Electromagnetic
  Theory}}}\ (\bibinfo  {publisher} {McGraw-Hill},\ \bibinfo {address} {New
  York},\ \bibinfo {year} {1941})\BibitemShut {NoStop}%
\bibitem [{\citenamefont {van~de Hulst}(1981)}]{Hulst:1981}%
  \BibitemOpen
  \bibfield  {author} {\bibinfo {author} {\bibfnamefont {H.}~\bibnamefont
  {van~de Hulst}},\ }\href@noop {} {\emph {\bibinfo {title} {Light Scattering
  by Small Particles}}}\ (\bibinfo  {publisher} {Dover},\ \bibinfo {address}
  {New York},\ \bibinfo {year} {1981})\BibitemShut {NoStop}%
\bibitem [{\citenamefont {Landau}\ \emph {et~al.}(1984)\citenamefont {Landau},
  \citenamefont {Lifshitz},\ and\ \citenamefont {Pitaevskii}}]{LanLifPit:1984}%
  \BibitemOpen
  \bibfield  {author} {\bibinfo {author} {\bibfnamefont {L.~D.}\ \bibnamefont
  {Landau}}, \bibinfo {author} {\bibfnamefont {E.}~\bibnamefont {Lifshitz}}, \
  and\ \bibinfo {author} {\bibfnamefont {L.}~\bibnamefont {Pitaevskii}},\
  }\href@noop {} {\emph {\bibinfo {title} {Electrodynamics of Continuous
  Media}}},\ \bibinfo {edition} {2nd}\ ed.\ (\bibinfo  {publisher} {Pergamon
  Press},\ \bibinfo {address} {Oxford},\ \bibinfo {year} {1984})\BibitemShut
  {NoStop}%
\bibitem [{Sta()}]{StabilityNote}%
  \BibitemOpen
  \href@noop {} {}\bibinfo {note} {A stability analysis was carried out in
  Ref.~\cite{SheltonBoninWalker:2005}, based on a first-order ODE introduced
  below, Eq.\ (\ref{eq:auxEOM}). The analysis here is based on the second-order
  ODE, Eq.\ (\ref{eq:auxAngleEOM}), which has the same stationary solutions but
  twice as many small-oscillation solutions around each stationary solution.
  The conclusions about stability or its absence are the same for the two
  ODEs.}\BibitemShut {Stop}%
\bibitem [{\citenamefont {Strogatz}(1994)}]{Strogatz:1994}%
  \BibitemOpen
  \bibfield  {author} {\bibinfo {author} {\bibfnamefont {Steven~H.}\
  \bibnamefont {Strogatz}},\ }\href@noop {} {\emph {\bibinfo {title} {Nonlinear
  Dynamics and Chaos with Applications to Physics, Biology, Chemistry, and
  Engineering}}}\ (\bibinfo  {publisher} {Addison-Wesley Publishing Company},\
  \bibinfo {address} {Reading},\ \bibinfo {year} {1994})\ \bibinfo {note} {,
  {S}ec. 4.3}\BibitemShut {NoStop}%
\bibitem [{\citenamefont {Adler}(1946)}]{Adler:1946}%
  \BibitemOpen
  \bibfield  {author} {\bibinfo {author} {\bibfnamefont {Robert}\ \bibnamefont
  {Adler}},\ }\bibfield  {title} {\enquote {\bibinfo {title} {A study of
  locking phenomena in oscillators},}\ }\href@noop {} {\bibfield  {journal}
  {\bibinfo  {journal} {Proceedings of the I.R.E}\ }\textbf {\bibinfo {volume}
  {34}},\ \bibinfo {pages} {351--357} (\bibinfo {year} {1946})}\BibitemShut
  {NoStop}%
\bibitem [{\citenamefont {Bender}\ and\ \citenamefont
  {Orszag}(1978)}]{BenderOrszag:1978}%
  \BibitemOpen
  \bibfield  {author} {\bibinfo {author} {\bibfnamefont {C.~M.}\ \bibnamefont
  {Bender}}\ and\ \bibinfo {author} {\bibfnamefont {S.~A.}\ \bibnamefont
  {Orszag}},\ }\href@noop {} {\emph {\bibinfo {title} {Advanced Mathematical
  Methods for Scientists and Engineers}}}\ (\bibinfo  {publisher}
  {McGraw-Hill},\ \bibinfo {address} {New York},\ \bibinfo {year}
  {1978})\BibitemShut {NoStop}%
\bibitem [{\citenamefont {Lichtenberg}\ and\ \citenamefont
  {Lieberman}(1983)}]{LichtLieb:1983}%
  \BibitemOpen
  \bibfield  {author} {\bibinfo {author} {\bibfnamefont {A.~J.}\ \bibnamefont
  {Lichtenberg}}\ and\ \bibinfo {author} {\bibfnamefont {M.~A.}\ \bibnamefont
  {Lieberman}},\ }\href@noop {} {\emph {\bibinfo {title} {Regular and
  Stochastic Motion}}}\ (\bibinfo  {publisher} {Springer Verlag},\ \bibinfo
  {address} {New York},\ \bibinfo {year} {1983})\BibitemShut {NoStop}%
\bibitem [{\citenamefont {Bonessi}\ \emph {et~al.}(2007)\citenamefont
  {Bonessi}, \citenamefont {Bonin},\ and\ \citenamefont
  {Walker}}]{Bonessi:2007}%
  \BibitemOpen
  \bibfield  {author} {\bibinfo {author} {\bibfnamefont {D.}~\bibnamefont
  {Bonessi}}, \bibinfo {author} {\bibfnamefont {K.}~\bibnamefont {Bonin}}, \
  and\ \bibinfo {author} {\bibfnamefont {T.}~\bibnamefont {Walker}},\
  }\bibfield  {title} {\enquote {\bibinfo {title} {Optical forces on particles
  of arbitrary shape and size},}\ }\href@noop {} {\bibfield  {journal}
  {\bibinfo  {journal} {Journal of Optics A: Pure and Applied Optics}\ }\textbf
  {\bibinfo {volume} {9}},\ \bibinfo {pages} {S228--S234} (\bibinfo {year}
  {2007})},\ \bibinfo {note} {{T}his issue of the journal is devoted to the
  topic of Optical Micromanipulation}\BibitemShut {NoStop}%
\bibitem [{\citenamefont {Liaw}\ \emph {et~al.}(2014)\citenamefont {Liaw},
  \citenamefont {Chen},\ and\ \citenamefont {Kuo}}]{LiawKuo:2014}%
  \BibitemOpen
  \bibfield  {author} {\bibinfo {author} {\bibfnamefont {J.~W.}\ \bibnamefont
  {Liaw}}, \bibinfo {author} {\bibfnamefont {Y.~S.}\ \bibnamefont {Chen}}, \
  and\ \bibinfo {author} {\bibfnamefont {M.~K.}\ \bibnamefont {Kuo}},\
  }\bibfield  {title} {\enquote {\bibinfo {title} {Rotating {A}u nanorod and
  nanowire driven by circularly polarized light},}\ }\href@noop {} {\bibfield
  {journal} {\bibinfo  {journal} {Optics Express}\ }\textbf {\bibinfo {volume}
  {22}},\ \bibinfo {pages} {26005--26015} (\bibinfo {year} {2014})}\BibitemShut
  {NoStop}%
\bibitem [{\citenamefont {Trojek}\ \emph {et~al.}(2012)\citenamefont {Trojek},
  \citenamefont {Chvatal},\ and\ \citenamefont {Zemanek}}]{TrojekZemanek:2012}%
  \BibitemOpen
  \bibfield  {author} {\bibinfo {author} {\bibfnamefont {J.}~\bibnamefont
  {Trojek}}, \bibinfo {author} {\bibfnamefont {L.}~\bibnamefont {Chvatal}}, \
  and\ \bibinfo {author} {\bibfnamefont {P.}~\bibnamefont {Zemanek}},\
  }\bibfield  {title} {\enquote {\bibinfo {title} {Optical alignment and
  confinement of an ellipsoidal nanorod in optical tweezers: a theoretical
  study},}\ }\href@noop {} {\bibfield  {journal} {\bibinfo  {journal} {Journal
  of the Optical Society of America: Optics and Image Science, and Vision}\
  }\textbf {\bibinfo {volume} {29}},\ \bibinfo {pages} {1224--1236} (\bibinfo
  {year} {2012})}\BibitemShut {NoStop}%
\bibitem [{\citenamefont {Horowitz}\ \emph {et~al.}(2011)\citenamefont
  {Horowitz}, \citenamefont {Alem\'{a}n}, \citenamefont {Christle},
  \citenamefont {Cleland},\ and\ \citenamefont
  {Awschalom}}]{HorowitzAwschalom:2012}%
  \BibitemOpen
  \bibfield  {author} {\bibinfo {author} {\bibfnamefont {V.~R.}\ \bibnamefont
  {Horowitz}}, \bibinfo {author} {\bibfnamefont {B.~J.}\ \bibnamefont
  {Alem\'{a}n}}, \bibinfo {author} {\bibfnamefont {D.~J.}\ \bibnamefont
  {Christle}}, \bibinfo {author} {\bibfnamefont {A.~N.}\ \bibnamefont
  {Cleland}}, \ and\ \bibinfo {author} {\bibfnamefont {D.~D.}\ \bibnamefont
  {Awschalom}},\ }\bibfield  {title} {\enquote {\bibinfo {title} {Electron spin
  resonance of nitrogen-vacancy centers in optically trapped nanodiamonds},}\
  }\href@noop {} {\bibfield  {journal} {\bibinfo  {journal} {Proceedings of the
  National Academy of Sciences}\ }\textbf {\bibinfo {volume} {109}},\ \bibinfo
  {pages} {13493--13497} (\bibinfo {year} {2011})}\BibitemShut {NoStop}%
\bibitem [{\citenamefont {Pedaci}\ \emph
  {et~al.}(2011{\natexlab{b}})\citenamefont {Pedaci}, \citenamefont {Huang},
  \citenamefont {van Oene}, \citenamefont {Barland},\ and\ \citenamefont
  {Dekker}}]{PedaciDekker:2011}%
  \BibitemOpen
  \bibfield  {author} {\bibinfo {author} {\bibfnamefont {F.}~\bibnamefont
  {Pedaci}}, \bibinfo {author} {\bibfnamefont {Z}~\bibnamefont {Huang}},
  \bibinfo {author} {\bibfnamefont {M.}~\bibnamefont {van Oene}}, \bibinfo
  {author} {\bibfnamefont {S.}~\bibnamefont {Barland}}, \ and\ \bibinfo
  {author} {\bibfnamefont {N.~H.}\ \bibnamefont {Dekker}},\ }\bibfield  {title}
  {\enquote {\bibinfo {title} {Excitable particle in an optical torque
  wrench},}\ }\href@noop {} {\bibfield  {journal} {\bibinfo  {journal} {Nature
  Physics}\ }\textbf {\bibinfo {volume} {7}},\ \bibinfo {pages} {259--264}
  (\bibinfo {year} {2011}{\natexlab{b}})}\BibitemShut {NoStop}%
\bibitem [{\citenamefont {Lipfert}\ \emph {et~al.}(2015)\citenamefont
  {Lipfert}, \citenamefont {van Oene}, \citenamefont {Lee}, \citenamefont
  {Pedaci},\ and\ \citenamefont {Dekker}}]{LipfertDekker:2015}%
  \BibitemOpen
  \bibfield  {author} {\bibinfo {author} {\bibfnamefont {J.}~\bibnamefont
  {Lipfert}}, \bibinfo {author} {\bibfnamefont {M.~M.}\ \bibnamefont {van
  Oene}}, \bibinfo {author} {\bibfnamefont {M.}~\bibnamefont {Lee}}, \bibinfo
  {author} {\bibfnamefont {F.}~\bibnamefont {Pedaci}}, \ and\ \bibinfo {author}
  {\bibfnamefont {N.~H.}\ \bibnamefont {Dekker}},\ }\bibfield  {title}
  {\enquote {\bibinfo {title} {Torque spectroscopy for the study of rotary
  motion in biological systems},}\ }\href@noop {} {\bibfield  {journal}
  {\bibinfo  {journal} {Chemical Reviews}\ }\textbf {\bibinfo {volume} {115}},\
  \bibinfo {pages} {1449--1474} (\bibinfo {year} {2015})}\BibitemShut {NoStop}%
\end{thebibliography}
\end{document}